\documentclass[acmsmall]{acmart}
\setcopyright{none}
\renewcommand\footnotetextcopyrightpermission[1]{}
\AtBeginDocument{%
  }

\begin{document}

\title{Making Sense of the Weather, Together: Collaborative Sensemaking in Severe Weather Livestreams}


\author{Julie A. Vera}
\affiliation{%
  \institution{University of Washington}
  \city{Seattle}
  \state{WA}
  \country{USA}
}
\email{jvera@uw.edu}
\renewcommand{\shortauthors}{Vera}

\author{Mark Zachry}
\affiliation{%
  \institution{University of Washington}
  \city{Seattle}
  \state{WA}
  \country{USA}
}
\author{David W. McDonald}
\affiliation{%
  \institution{University of Washington}
  \city{Seattle}
  \state{WA}
  \country{USA}
}

\begin{abstract}
This paper examines collaborative sensemaking during severe weather events through the emerging phenomenon of "weatherfluencers" or content creators who livestream meteorological interpretation on platforms like YouTube. Drawing from sensemaking theory, crisis informatics, and platform studies, we analyze how these creators navigate the sociotechnical dynamics of interpreting severe weather in real time with distributed audiences. Through critical incident analysis of 13 Particularly Dangerous Situation (PDS) storm warnings across three prominent weatherfluencers, we identify three key practices: multi-source information triangulation, temporal bridging techniques, and platform-specific adaptations that transform entertainment interfaces into safety-critical communication channels. Our analysis shows how these practices challenge existing models of crisis communication by integrating distributed expertise, collapsing temporal frames, and reconfiguring platform affordances. This research contributes to understanding how informal emergency communicators mediate between institutional alerting systems and public needs, and how visual, multimodal crisis communication differs from text-centered approaches.
\end{abstract}

\begin{center}
\textit{Accepted to CSCW 2026. This arXiv version is the author-accepted manuscript.}
\end{center}




\maketitle

\section{Introduction}
During severe weather events, livestreaming platforms have emerged as critical channels for real-time information dissemination. A growing community of content creators, whom we term "weatherfluencers," have developed specialized practices for interpreting complex meteorological data for public audiences. Drawing from Hurcombe’s concept of "newsfluencers" \cite{hurcombe2024conceptualising}, these creators adapt influencer logics to weather communication while navigating tensions between entertainment imperatives and public safety responsibilities.

Weatherfluencers operate at a distinctive intersection of entertainment, education, and public safety, translating technical data into accessible interpretations while enabling collaborative sensemaking through platform affordances. Unlike traditional TV meteorologists constrained by broadcast formats, these creators leverage digital platforms to coordinate distributed expertise by triangulating multiple information sources, engaging audiences in real-time information verification, and performing transparent analytical processes that transform passive viewers into active participants in collective sensemaking.

This emergence occurs amid broader shifts in crisis communication, including diminished trust in governmental institutions \cite{bell_public_2025} and increasing reliance on informal digital intermediaries for safety-critical information \cite{tomczyk_sharing_2025, widjaya_closer_2025}. As traditional weather infrastructure faces resource constraints \cite{oklahoma_as_2025, dance_national_2025}, weatherfluencers fill informational gaps by positioning themselves as intermediaries between institutional alerting systems and public understanding, as publics increasingly turn to platforms for up-to-the-minute information during unfolding events.

However, significant gaps remain in our understanding of how live, interactive video platforms function as collaborative sensemaking systems during high-stakes crisis events. Prior crisis informatics research has largely focused on asynchronous, text-based social media, such as Twitter and Facebook, leaving real-time, multimodal sensemaking practices comparatively underexplored \cite{starbird2014rumors, palen2016crisis}. Similarly, while sensemaking theory has emphasized retrospective interpretation, severe weather livestreams require participants to construct meaning prospectively as threats unfold in real time \cite{weick1995sensemaking}. As a result, existing frameworks offer limited explanation of how interpretation, credibility, and coordination unfold in real time during unfolding crises.

To address these gaps, this study examines how weatherfluencers and their audiences collaboratively make sense of severe weather events through three research questions: (1) What collaborative sensemaking practices emerge between weatherfluencers and their audiences during critical safety moments? (2) How do weatherfluencers engage in temporal sensemaking during severe weather events? (3) How do platform affordances and constraints shape weatherfluencers' communication practices during safety-critical moments?

Through critical incident analysis of 13 Particularly Dangerous Situation (PDS) warnings across the livestreams of three prominent weatherfluencers, we identify practices that transform entertainment-oriented livestreaming platforms into sites of safety-critical interpretation. This study contributes to research in computer-supported cooperative work and crisis informatics by documenting time-sensitive, collaborative sensemaking, analyzing how platform affordances shape emergency communication, and extending crisis informatics beyond asynchronous, text-centered models to account for real-time, visual, and publicly performed interpretation.

\section{Literature Review}
This section situates our study within research on crisis informatics, collaborative sensemaking, participatory media, credibility in weather communication, and platform affordances. Collectively, these literatures explain how information circulates, how meaning is constructed under uncertainty, and how sociotechnical conditions shape interpretation during emergencies. We synthesize this work to motivate an analysis of real-time, platform-mediated crisis sensemaking.

\subsection{Crisis Informatics and Social Media}
Crisis informatics examines the role of information and communication technologies during emergency events, with particular attention to how digital platforms reshape coordination, information flow, and public participation during disasters \cite{hughes2015social,palen2016crisis}. Early work by Palen and colleagues established that social media platforms enable decentralized crisis communication, allowing affected individuals and volunteers to contribute information, interpretation, and coordination efforts previously dominated by institutional actors  \cite{palen2009crisis, palen2007citizen}.

As social media platforms have become embedded in emergency response, they have increasingly functioned as critical information infrastructure rather than auxiliary communication channels \cite{sutton2014warning, murthy2017social, gui2017managing, kaufhold2024exploring}. Empirical studies demonstrate how platforms such as Twitter/X support situational awareness, rapid information dissemination, and collective coordination during disasters \cite{vieweg2010microblogging, sutton2014warning, reuter2018fifteen}. More recent work has examined how digital volunteers and informal participants contribute to crisis response through practices such as information amplification, verification, and translation \cite{dailey2017social}.

A central concern in crisis informatics research involves the relationship between institutional and informal crisis communicators. Prior studies document persistent challenges related to information verification, authority conflicts, and the spread of misinformation when official and unofficial sources coexist during emergencies \cite{starbird2014rumors, crawford2015limits, bunker2020convergence}. In response, scholars have highlighted the need to better account for the interaction and coordination between institutional and informal communication practices, recognizing that these actors often play complementary roles during crisis response \cite{hughes2015social}.

However, much of this literature conceptualizes informal participants as contributors of reports, signals, or verification labor rather than as sustained interpreters of unfolding events. Crisis informatics research has paid comparatively less attention to how non-institutional actors assume interpretive roles in real time, publicly synthesizing information, contextualizing risk, and guiding audience understanding as events unfold. This gap motivates closer examination of weatherfluencers as emergent crisis actors who operate at the intersection of institutional alerts, platform infrastructures, and public sensemaking during severe weather livestreams. These emergent interpretive roles parallel broader shifts in participatory media, where non-institutional creators increasingly act as public-facing interpreters of complex events, reshaping how publics encounter and make sense of unfolding crises.

\subsection{Collaborative Sensemaking and Temporal Dynamics in Crisis}
While crisis informatics research examines who participates in crisis communication and how information circulates across platforms, sensemaking scholarship provides an analytic lens for understanding how meaning is constructed under conditions of uncertainty. Weick’s work on sensemaking highlights how crises disrupt taken-for-granted structures of meaning, producing "cosmology episodes" in which actors struggle to re-establish coherence amid uncertainty and breakdown \cite{weick_collapse_1993, weick_reflections_2010, weick_managing_2015}. Crisis situations thus foreground sensemaking processes by introducing ambiguity and urgency that destabilize established interpretive frames and demand active meaning-making.

Research on collaborative sensemaking emphasizes that interpretation is not an individual cognitive process but a socially distributed activity shaped through interactions between individuals. Distributed cognition frameworks demonstrate how knowledge is coordinated across networks of people, artifacts, and technologies rather than residing within any single actor \cite{hutchins1995cognition, hollan2000distributed}. In crisis contexts, this distribution of expertise often extends beyond formal organizational boundaries to include volunteers, affected communities, and informal participants \cite{dynes1970organized, majchrzak2007coordinating}. Prior work documents how online crowds engage in emergent problem-solving, information synthesis, and coordination during disasters \cite{starbird2011voluntweeters}.

Sensemaking in crisis contexts is often accomplished through ongoing interaction rather than through stable representations alone. Prior research on interaction and collaborative work demonstrates how meaning is produced through talk, gesture, and reference to shared visual or material resources, particularly in situations requiring rapid coordination and interpretation \cite{goodwin_action_2000, hutchins1995cognition}. In digitally mediated environments, this interactional work becomes visible through continuous explanation, questioning, and revision as participants respond to unfolding information \cite{weick1995sensemaking, starbird2011voluntweeters}. In livestreamed crisis communication, sensemaking unfolds in real time, with interpretation, anticipation, and coordination occurring simultaneously rather than sequentially, foregrounding temporality as a central feature of collaborative meaning-making \cite{mirbabaie2020social, reuter2018fifteen}.

The temporal dimension of sensemaking introduces additional complexity in crisis contexts. Although sensemaking has often been characterized as retrospective \cite{weick1995sensemaking}, subsequent work has emphasized its ongoing and prospective dimensions, particularly in crisis contexts where actors must interpret unfolding situations and anticipate future consequences \cite{maitlis2010sensemaking, gephart2010future}. Crisis response involves coordinating understanding across past experience, present conditions, and anticipated developments \cite{kaplan2013temporal}. Studies of emergency response describe how teams manage multiple temporal horizons through practices such as temporal trajectory alignment \cite{reddy2006temporality}, while uncertainty about timing significantly shapes crisis decision-making \cite{liu2016communicating}.

Platform-mediated communication further reshapes these temporal dynamics. Social media timelines collapse retrospective accounts, real-time interpretation, and forward-looking speculation into shared streams, altering how narratives are constructed and updated during crises \cite{reuter2018fifteen}. Mirbabaie et al. \cite{mirbabaie2020social} describe this phenomenon as temporal compression, in which multiple temporal orientations coexist within the same information space. Livestreaming intensifies this condition by producing continuous, multimodal broadcasts where interpretation, coordination, and anticipation occur simultaneously rather than sequentially.

Despite extensive scholarship on sensemaking and crisis response, prior work has largely examined sensemaking within organizations or formal response teams, and has often emphasized retrospective, asynchronous accounts of crisis communication. As a result, less is known about how collaborative sensemaking unfolds publicly and in real time on livestreaming platforms, where heterogeneous audiences and non-institutional actors must interpret unfolding events and anticipate consequences simultaneously. This gap motivates closer examination of severe weather livestreams as sites of collective, temporally entangled sensemaking, where meaning, anticipation, and action are co-constructed through ongoing interaction.

\subsection{Newsfluencers and Participatory Media}
The emergence of "newsfluencers" reflects a broader shift toward creator-led news production within platformized media ecosystems, where individual content creators adapt influencer practices to the dissemination and interpretation of news \cite{abidin2021networked, lewis2018decade}. Hurcombe \cite{hurcombe2024conceptualising} defines newsfluencers as "platformatised creators who operate according to the economic and cultural logics of online influencers to produce news content for participatory audiences." This framework provides a useful point of departure for understanding weatherfluencers as a specialized subset of creator-led information actors.

Newsfluencer scholarship highlights how creators operate at the intersection of platform infrastructures, entrepreneurial labor, and audience relationships. Hurcombe \cite{hurcombe2024conceptualising} identifies recurring dimensions of newsfluencer practice, including reliance on platform monetization systems, audience-supported business models, relational labor centered on engagement and responsiveness, and participatory cultures shaped by parasocial interaction. This work illustrates how informational authority in digital environments is increasingly performed through ongoing interaction rather than institutional affiliation alone.

These dynamics build on longer-standing traditions of participatory journalism, in which publics play active roles in news production, interpretation, and circulation \cite{coddington2015clarifying, lewis2015actors, singer2011participatory, wall2015citizen}. However, research on newsfluencers and participatory journalism has largely examined contexts in which informational stakes are moderate and interpretive timelines are relatively flexible, such as political commentary, social issues, or explanatory reporting. As a result, less attention has been paid to how creator-led information practices operate under conditions of acute risk and temporal pressure.

Weather communication presents a distinctive and analytically consequential case. Prior research on broadcast meteorology shows that public interpretation of weather forecasts is shaped by uncertainty communication, trust in presenters, and perceptions of expertise \cite{demuth2012creation, sherman2013public}. Studies of traditional broadcast meteorology emphasize institutional credibility and professional norms as foundations for effective warning communication \cite{henson2013weather, drost2016severe}. Weatherfluencers diverge from this model by operating outside formal meteorological institutions, sometimes without formal credentialing, while nonetheless providing real-time interpretation of complex, safety-critical data to large audiences.

While influencer and newsfluencer research helps explain how creators cultivate engagement and perform interpretive authority, it offers limited insight into how these practices operate when audiences may rely on creator interpretations to make time-sensitive safety decisions. Severe weather livestreams place creators in a qualitatively different role: weatherfluencers must simultaneously sustain attention within platformized media environments and publicly interpret evolving, safety-critical data in real time. In these contexts, engagement is not merely a commercial or relational concern, but a condition for effective sensemaking and potential protective action. This study extends newsfluencer and participatory media scholarship by examining how creator-led interpretive practices are reconfigured in severe weather livestreams, where sensemaking, credibility, and action become tightly coupled under conditions of real-time uncertainty.

\subsection{Credibility and Trust in Weather Communication}

Weather communication is a domain in which credibility has long been treated as essential \cite{mileti_communication_1990, ripberger_false_2015}. Severe weather warnings involve high stakes and asymmetric access to expertise, requiring publics to act on information they cannot independently verify. Decades of research show that perceived credibility of weather information sources shapes how warnings are interpreted and whether protective action is taken \cite{mileti_communication_1990, ripberger_false_2015, burgeno_impact_2020, burgeno_impact_2023}. Consequently, prior work has treated credible sources as a condition for effective weather communication, particularly in high-stakes, safety-critical contexts.

In traditional meteorological contexts, credibility is closely tied to institutional affiliation, professional credentials, and broadcast media infrastructure. Empirical studies show that audiences place high levels of confidence in local television meteorologists, who are perceived as authoritative, familiar, and reliable sources of weather information \cite{bloodhart_local_2015}. Within this model, trust and credibility function as a relatively stable condition that supports warning effectiveness, and is commonly conceptualized as an outcome or mediating variable linking warning characteristics and behavioral response \cite{ripberger_false_2015, trainor_tornadoes_2015, losee_need_2018, weyrich_dealing_2019}.

Digital and platform-mediated environments complicate this model of credibility. Research in human-computer interaction demonstrates that credibility online is shaped not only by source identity, but also by interface cues, communicative performance, and interactive signals \cite{flanagin_role_2007, fogg_elements_1999, metzger_making_2007}. Recent research on social media influencers shows that audiences respond to authenticity, transparency, and direct engagement as signals of credibility, even in the absence of any formal expertise. Work in digital marketing and communication finds that strategies such as information disclosure and expressive authenticity increase audience engagement and perceived credibility, suggesting that interactive, relational cues can matter just as much or more than formal authority cues \cite{thi_vi_influencer_2025, khalfallah_authenticity_2025}.

In crisis contexts, credibility takes on heightened importance due to temporal pressure, epistemic uncertainty, and the consequences of delayed or incorrect interpretation. Prior crisis informatics research has documented how misinformation, verification challenges, and authority disputes emerge when institutional and informal sources coexist during emergencies \cite{starbird2014rumors, hughes2015social, palen2007citizen}. However, this work has largely examined credibility through post-hoc analyses or asynchronous communication, offering limited insight into how credibility is enacted and sustained in interaction during live, unfolding events.

This gap is particularly salient in the context of severe weather livestreams, where non-institutional actors publicly interpret evolving data for heterogeneous audiences in real time. Rather than evaluating trust as an outcome or benchmarking accuracy against external ground truth, this study examines how credibility becomes sufficient for collective sensemaking to proceed under conditions of uncertainty. By focusing on interactive practices such as transparency, source triangulation, and responsiveness, the study addresses how credibility is made visible, negotiated, and repaired during livestreamed crisis communication, and how these practices shape the presentation and interpretation of information about storm severity, trajectory, and appropriate response. However, prior work has not examined how credibility is made visible and actionable in real-time, public-facing livestreams during active crises, which this study addresses by analyzing observable practices under conditions of temporal pressure and epistemic uncertainty.

\subsection{Platform Affordances and Livestreamed Crisis Communication}
Research on platform affordances emphasizes how technical features shape the possibilities for communication, coordination, and interpretation within sociotechnical systems \cite{norman1999affordance, evans2017explicating, bucher2018affordances}. In crisis contexts, these affordances influence not only how information circulates, but how uncertainty, risk, and urgency are made visible and interpretable under time pressure \cite{reuter2011social}. Prior work has examined how features such as character limits, hashtagging, and threading structure crisis communication on platforms including Twitter/X, social networking sites, and Reddit \cite{bruns2014crisis, white2009online, leavitt2017role}.

Beyond individual features, platform studies highlight how crisis communication unfolds within broader logics of governance, visibility, and monetization. Civic information infrastructures increasingly rely on privately governed platforms whose algorithmic systems and engagement-driven design priorities shape which information gains prominence during emergencies \cite{gillespie2010politics, van2013understanding}. These dynamics can introduce tensions between visibility, accuracy, and public safety, particularly when engagement metrics influence information prominence during high-stakes events.

At the same time, platforms are not used as designed. Research on infrastructuring and appropriation demonstrates how users adapt and reconfigure platforms to meet emergent needs, including in high-stakes and time-sensitive situations \cite{pipek_infrastructuring_2009, star_infrastructure_2002, lingel2017digital}. In crisis contexts, such adaptations often involve assembling visual, textual, and interactive resources to support interpretation and coordination under conditions of uncertainty.

Visual affordances are especially consequential for weather communication, where understanding depends on interpreting spatial and temporal representations such as radar imagery, maps, and warnings. Prior research shows that visual artifacts can support shared situational awareness by providing common reference points through which participants interpret evolving conditions \cite{thelwall2007ruok, liu2008search, reuter2018fifteen}. However, much of this work has examined visual sensemaking in asynchronous or tool-centered contexts rather than in live, publicly interactive environments.

Livestreaming platforms bring these affordances together within a single real-time communicative setting that integrates continuous broadcasting, persistent visual displays, and audience interaction. While prior research has documented practices such as continuous narration and direct audience engagement in livestreamed communication \cite{dougherty2011live, hamilton2014streaming}, crisis informatics research has largely focused on asynchronous or text-based platforms. As a result, less is known about how livestream affordances shape interpretation and coordination as crisis events unfold.

We address this gap by examining YouTube livestreams as sociotechnical environments in which platform affordances actively structure collaborative sensemaking during severe weather events. Instead of treating platforms as neutral channels for information dissemination or evaluating trust against external ground truth, we analyze how livestream affordances enable weatherfluencers and audiences to interpret evolving data, negotiate uncertainty, and coordinate understanding in real time. In doing so, the study clarifies how platform design shapes the practices through which sensemaking proceeds under conditions of temporal pressure and informational ambiguity.

\section{Research Questions}

Building on the literature reviewed above and addressing the identified gaps in understanding collaborative sensemaking in livestreamed crisis contexts, we pose the following research questions:

\begin{enumerate}
    \item What collaborative sensemaking practices emerge between weatherfluencers and their audiences during critical safety moments in severe weather livestreams?
    \item How do weatherfluencers engage in temporal sensemaking during severe weather events?
    \item How do platform affordances and constraints of livestreaming services shape weatherfluencers' communication practices during safety-critical moments?
\end{enumerate}

\section{Positionality Statement}
The first author is a National Weather Service SkyWarn Storm Spotter with both technical familiarity and an ethical commitment to public safety during severe weather. This perspective informs our analysis in two key ways. First, our meteorological knowledge enables recognition of nuanced technical discussions and accuracy assessment that might elude researchers without domain expertise. Second, it shapes our understanding of the ethical dimensions that weatherfluencers navigate when communicating potentially life-saving information.

We acknowledge that this technical background may create analytical blind spots; for example, what we find technically accurate or important may differ from what general audiences prioritize. Throughout this research, we balance our technical perspective with careful attention to how non-expert viewers engage with content, being mindful not to overprivilege technical accuracy at the expense of understanding broader communication patterns. 

\section{Methods}
\begin{figure}
    \centering
    \includegraphics[width=1\linewidth]{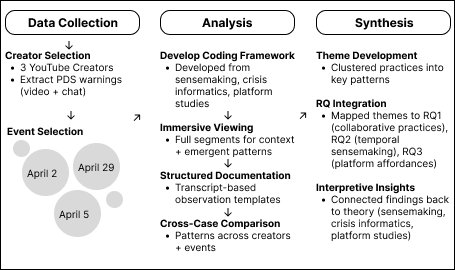}
    \caption{Multi-step data collection and analysis pipeline}
    \Description{Figure showing the workflow from data collection through analysis through data synthesis.}
    \label{fig:comprehensive-flow}
\end{figure}

This study examines how weatherfluencers and their audiences collaboratively make sense of severe weather events through platform-mediated interaction. We employ the critical incident technique to analyze Particularly Dangerous Situation (PDS) warnings issued during live events as moments when routine weather monitoring transforms into safety-critical communication. Figure~\ref{fig:comprehensive-flow} illustrates our four-stage methodological pipeline: creator and event selection, incident-centered data collection, collaborative coding, and cross-case analysis.

\subsection{Creator and Event Selection}
We systematically selected livestream creators based on four criteria aligned with our research questions: (1) regular coverage of severe weather events; (2) use of the "at-the-desk" livestreaming format focusing on radar interpretation; (3) real-time public-facing meteorological analysis; and (4) substantial audience engagement of at least 1,000 concurrent viewers per stream. We consulted Reddit discussions and Discord communities dedicated to severe weather education to identify candidates meeting these criteria.

\subsubsection{Creator Profiles}
Our final dataset comprises three YouTube-based livestreams from prominent weatherfluencers covering major tornado outbreaks in Spring 2025. While this represents a small sample (N=3), it captures a concentrated set of sustained, high-visibility livestreaming practices within a specialized sub-niche of severe weather content on YouTube. Weather-related content on YouTube is extensive and heterogeneous, encompassing routine forecast updates, model discussions, post-event analysis, and institutionally-produced coverage by media organizations. In contrast, the creators examined here are distinguished by their emphasis on multi-hour, multi-location, real-time livestreaming during active severe weather events, integrating radar interpretation, audience interaction, and rapid updates as events unfold. This focus situates them within a platform-native, interpretive practice that differs from both forecast-centric channels and institutional weather media. Creator A (89.1K subscribers) comprises a main host, two formally trained consulting meteorologists, and several storm chasers. Creator B (1.19M subscribers) is an enthusiast and student who largely reports solo. Creator C (2.7M subscribers) comprises a main host and at least one formally trained meteorologist. Accordingly, the creators examined here should be understood as illustrative of this specific, platform-native practice of live severe weather interpretation, rather than as representative of all weatherfluencer activity on YouTube. Table~\ref{tab:Creators} summarizes the creator profiles.
\begin{table}
\centering
\begin{tabular}{|l|l|l|l|l|}
\hline
\textbf{Channel} & \textbf{Joined YouTube} & \textbf{No. Videos} & \textbf{No. Views} & \textbf{No. Subscribers\footnotemark} \\
\hline
Creator A & Sep 28, 2022 & 945 & 13.7M & 89.1K \\
\hline
Creator B & Aug 31, 2014\footnote{This creator's weather-related videos begin in September 2021} & 1,568 &200M &1.19M \\
\hline
Creator C &Dec 5, 2012 &593 &403M &2.7M \\
\hline
\end{tabular}
\caption{Creator profiles: Date the creator joined YouTube, number of videos, number of total views, and total number of subscribers}
\label{tab:Creators}
\end{table}
\footnotetext{As of May 7, 2025}

\subsubsection{Event Profiles}
The April 2-5, 2025 and April 29, 2025 tornado outbreaks received a notable amount of attention from the National Weather Service, traditional media outlets, and weatherfluencers due to their timing \footnote{April is typically the beginning of an active tornado season in North America.} and perceived threat to a large portion of the central and southern United States. These events affected large portions of the central and southern United States during peak tornado season, allowing us to compare coverage across multiple creators and geographical areas while maintaining comparable levels of threat urgency. Table \ref{tab:event_sampling} summarizes the specific livestreams analyzed. In total, we analyzed one multi-hour severe weather event for each creator, with event lengths ranging from 5.5 to 12 hours. We analyze a single severe weather event per creator to enable close comparison of sensemaking practices under broadly comparable conditions. Each selected event involved many hours of sustained livestream coverage, providing substantial temporal depth to observe recurring interaction patterns, shifts in informational availability, and changes in audience participation over time, enabling observation of sensemaking practices across extended temporal spans within a single event.

\begin{table}[htbp]
  \centering
  \begin{tabular}{|c|c|c|p{3.5cm}|}
    \hline
    \textbf{Creator} & \textbf{Stream Date} & \textbf{Total Length (hrs)} & \textbf{Location/Area of Focus} \\
    \hline
    Creator A & April 2, 2025 & 5.5 & Plains States (OK, TX, AR)\\
    \hline
    Creator B & April 3, 2025 & 12 & Deep Southern States (MS, TN, AR) \\
    \hline
    Creator C & April 29, 2025 & 12 & Deep Southern States (LA, MS, TN, AR) \\
    \hline
  \end{tabular}
  \caption{Events Sampled, Length, and Geographic Focus}\label{tab:event_sampling}
\end{table}
\subsection{Critical Incident Sampling Strategy}
\begin{figure}
    \centering
    \includegraphics[width=1\linewidth]{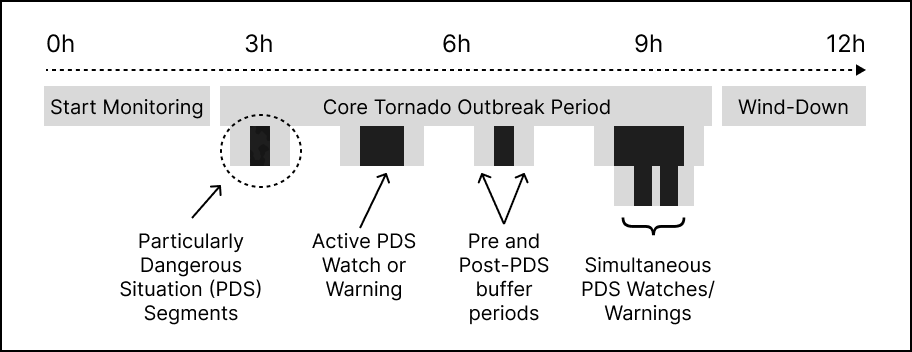}
    \caption{Critical Incident Sampling and PDS Watch/Warning Structure}
    \Description{Image demonstrating the multi-step collection and data analysis pipeline, along with the structure of PDS warnings within the greater flow of the livestream.}
    \label{fig:sampling-map}
\end{figure}

We employ the critical incident technique \cite{flanagan1954critical, butterfield2005fifty} to analyze moments when weatherfluencers shift from routine monitoring to high-stakes information interpretation. Given the extended duration of severe weather livestreams, often approaching 12 hours, we implemented a purposeful, incident-centered sampling strategy focused on Particularly Dangerous Situation (PDS) watches and warnings as critical incidents. PDS watches and warnings are rare, elevated alerts representing the second-highest level of severe weather alert in the US system, activated only when National Weather Service forecasters observe conditions likely to produce significant danger to life and property \cite{NOAA2025PDS}. Figure \ref{fig:sampling-map} illustrates the simplified temporal structure through which PDS warnings typically emerge within livestreams, situating them within periods of early-stream storm monitoring, periods of active warnings, buffer time before and after warnings, and occasional overlap across multiple PDS events.

PDS warnings function similarly to what Weick \cite{weick_cosmos_1997} terms "cosmology episodes" where both sense and the means of sensemaking collapse simultaneously. They create ideal analytical windows for three theoretically-grounded reasons: they generate temporal compression requiring integration of past, present and future frames \cite{reddy_temporality_2006}; they produce epistemic uncertainty requiring collaborative validation \cite{fischer2015building}; and they necessitate distributed expertise across creator, audience, and institutional sources \cite{majchrzak2007coordinating}.

For each identified PDS warning, we extracted a 10-minute segment: 5 minutes before the warning was issued, the warning duration, and 5 minutes after cancellation. In cases of overlapping warnings, we sampled content from 5 minutes before the first PDS warning and 5 minutes after the cancellation of the last-announced PDS warning. This bracketing allowed us to analyze how creators and audiences made sense of meteorological data in anticipation of, during, and after safety-critical events. A total of 13 PDS tornado warnings were identified across the three sampled streams during the April 2--3, 2025 and April 29, 2025 tornado outbreaks, resulting in 453 minutes of analyzed content. All segments were annotated using YouTube Transcripts and Live Chat Replay, with approximately 300 chat messages analyzed per stream during PDS-warned events. Table \ref{tab:pds_warnings} summarizes the number of PDS warnings and total PDS-focused analysis time across creators. Given the high volume of chat activity (often tens of thousands of messages in a 12-hour stream), we employed purposeful sampling, analyzing approximately 300 messages per stream during PDS-warned events. Chat messages were analyzed using the same coding framework applied to creator communications, outlined in detail in \ref{Coding Framework and Analytical Approach}. 

We analyze this activity through a \textbf{focused content analysis} grounded in theories of sensemaking \cite{weick1995sensemaking}, concepts of newsfluencers \cite{hurcombe2024conceptualising}, and platform studies \cite{van2013understanding, gillespie2010politics}. To support this analysis, we draw conceptually from the trace ethnography lens \cite{geiger2011trace}, which attends to how digital traces, such as chat logs, creator commentary, and on-screen visuals, reflect sociotechnical practices. While we do not conduct full ethnographic immersion, this perspective helps us interpret how participants navigate uncertainty and urgency through the digital traces they leave behind.

\begin{table}[htbp]
  \centering
  \begin{tabular}{|c|c|c|p{2.5cm}|}
    \hline
    \textbf{Creator} & \textbf{Event Date} & \textbf{PDS Warnings} & \textbf{Total PDS Time Analyzed}\\
    \hline
    Creator A& April 2, 2025& 1 & 73 minutes \\
    \hline
    Creator B& April 3, 2025& 8 & 200 minutes \\
    \hline
    Creator C& April 29, 2025& 4 & 180 minutes \\
    \hline
    \hline
    \textbf{Total} & 3 Events & 13 & 453 minutes \\
    \hline
  \end{tabular}
  \caption{Summary of PDS Warnings Observed Across Livestreams}
  \label{tab:pds_warnings}
\end{table}

\subsection{Coding Framework and Analytical Approach}
\label{Coding Framework and Analytical Approach}
We developed a lightweight qualitative coding framework grounded in three theoretical traditions: sensemaking theory and principles \cite{weick1995sensemaking, weick2005organizing}, crisis informatics \cite{palen2016crisis, palen2007crisis}, and platform studies \cite{gillespie2010politics, van2013understanding}. This interpretive approach aligns with research traditions examining emergent digital practices \cite{geiger2011trace}, where rich descriptions of communication patterns reveal more insight than quantitative metrics in novel contexts.

The coding framework organized observations into four core domains aligned with our research questions:

\begin{itemize}
    \item \textbf{Collective Sensemaking}: Practices where creators and audiences co-construct understanding, including using information triangulation and query-driven redirection.
    \item \textbf{Sensemaking}: Creator-led interpretive strategies, including information scaffolding (connecting technical data to familiar concepts) and transparency in analytical processes (breadcrumbing familiar concepts to explain something more complex)
    \item \textbf{Temporal Orientation}: How creators manage and navigate time as severe weather unfolds, particularly to help make sense of a situation, including adding retrospective framing ("This is what we saw in Lake City earlier today"), real-time interpretation, and projections of future threats.
    \item \textbf{Platform Affordances and Constraints}: How YouTube's features shape communication, including how visual tools are presented, and how any platform limitations are handled. 
\end{itemize}

While these domains were informed by established theoretical traditions, the specific codes within each domain were developed through an abductive analytical process that integrated theory with close, iterative engagement with the data. Rather than applying a fully deductive codebook from prior literature, we began with theory-informed sensitizing concepts and refined them through repeated viewing of livestream segments and associated chat interactions during PDS warnings.

The appendix summarizes this coding schema, illustrating how high-level analytic domains informed by theory were operationalized through empirically observed communicative practices. For example, sensemaking theory motivated the broader Sensemaking domain, while specific practices such as Knowledge Scaffolding and Transparency emerged inductively from observing how creators explained complex meteorological data in real time. In this exploratory investigation, we prioritized identifying recurring patterns and distinctive practices across creators rather than quantifying code frequencies.

\subsection{Integrated Analytical Process}
Our analytical process prioritizes rich description and theoretical insight over systematic coding. Rather than attempting to quantify weatherfluencer practices, we sought to develop thick descriptions that capture the complex interplay between temporal dynamics, collaborative sensemaking, and platform affordances during critical weather moments. The analysis unfolds in three interconnected phases:

\subsubsection{Phase 1: Immersive Engagement with Stream Segments}
We begin by viewing each livestream segment in its entirety, immersing ourselves in the contextual flow of the coverage. This approach acknowledges the integrated nature of weatherfluencer practices, where temporality, collaboration, and platform features operate simultaneously rather than as separate dimensions. During this initial viewing, we noted significant moments where multiple elements converged, such as when NWS warnings were announced, attention shifted to different areas on radar maps, and chat activity intensified. We also paid particular attention to patterns in chat activity, including spikes in message frequency during key announcements, information sharing from viewers in affected areas, recurring questions that prompted creator responses, and moderator interventions that shaped information flow. Participation was uneven, with a relatively small subset of viewers contributing repeatedly while many others remained silent, reflecting the self-selected nature of synchronous engagement in livestream environments. Our analysis treats these interactions as evidence of how audience participation is made visible and actionable within livestreams, rather than as evidence of how information is perceived, trusted, or acted upon by viewers.

\subsubsection{Phase 2: Structured Documentation with Theoretical Linkages}
After identifying the timestamps of PDS warnings, we returned to each segment for systematic documentation. Using YouTube's Transcript feature, we created structured observation records capturing the meteorological context, creator's sensemaking techniques, audience interactions, and visual tools used to communicate risk. Appendix~\ref{appendix:analysis-template} contains a copy of our observation template. This structured documentation phase served as the foundation for our cross-case analysis, providing rich descriptive data grounded in theory.

\subsubsection{Phase 3: Cross-Case Pattern Recognition and Theoretical Integration}
The final analytical phase compared cases across creators and warning events to identify patterns and emergent themes. We developed analytical memos grounding these themes in specific examples while connecting to our theoretical framework. This comparative process included cross-creator analysis, cross-event analysis, and pattern identification across communication strategies, audience dynamics, and platform mediations.

\section{Findings}

Rather than simply relaying institutional alerts, weatherfluencers orchestrate networked crisis expertise through multi-source information synthesis, create public temporal sensemaking infrastructure that collapses past, present, and future into coherent narratives, and engage in platform appropriation that transforms entertainment affordances into civic safety systems. These practices challenge existing models of crisis communication by distributing expertise across human-technical networks while maintaining coherence through real-time collaborative interpretation. We use "collaborative" to describe the interactive processes through which creators solicit, respond to, and incorporate audience contributions in real time, rather than to imply shared understanding, consensus, or downstream impact. Table \ref{tab:findings-RQ-map} maps our key findings to our research questions. Our analysis focuses on observable interactional practices through which interpretation is made visible, rather than on how audiences ultimately perceive or act on this information.

\begin{table}[htbp]
\centering
\small % Makes the text smaller to fit better
\begin{tabular}{|p{2.5cm}|p{2.2cm}|p{3.5cm}|p{3.8cm}|}
\hline
\textbf{Research Question} & \textbf{Primary Theme} & \textbf{Key Practices} & \textbf{Theoretical Contribution} \\
\hline
\textbf{RQ1:} What collaborative sensemaking practices emerge? & 
Collaborative Sensemaking Practices & 
$\bullet$ Multi-source triangulation \newline
$\bullet$ Audience engagement \newline
$\bullet$ Technical translation \newline
$\bullet$ Distributed warning networks \newline
$\bullet$ Transparent analysis & 
\textbf{Networked Crisis Expertise:} Authority is made visible through the orchestration of multiple information streams, not necessarily credentials \\
\hline
\textbf{RQ2:} How do weatherfluencers engage in temporal sensemaking? & 
Temporal Communication Strategies & 
$\bullet$ Historical referencing \newline
$\bullet$ Real-time interpretation \newline
$\bullet$ Anticipatory projection \newline
$\bullet$ Temporal integration \newline
$\bullet$ Explicit timing limitations & 
\textbf{Public Temporal Sensemaking Infrastructure:} Platforms become tools for navigating past, present, and future simultaneously \\
\hline
\textbf{RQ3:} How do platform affordances shape communication? & 
Platform-Specific Adaptations & 
$\bullet$ Multi-window displays \newline
$\bullet$ Visual annotations \newline
$\bullet$ Technical workarounds \newline
$\bullet$ Attention management & 
\textbf{Platform Appropriation:} Entertainment infrastructures transformed into civic safety systems through creative adaptation \\
\hline
\end{tabular}
\caption{Summary of Research Questions, Practices, and Theoretical Contributions}
\label{tab:findings-RQ-map}
\end{table}
\subsection{Sensemaking Practices}
During critical safety moments, weatherfluencers orchestrate distributed sensemaking through five recurring practices that convert livestreams from one-way commentary into interactive, multi-source situational awareness. Across creators, credibility was not asserted through credentials alone; instead, it was performed through visible cross-checking, deliberate solicitation of audience input, accessible translation of technical indicators, and real-time reasoning that made uncertainty legible.

\subsubsection{Visible Triangulation as Epistemic Authority}
Weatherfluencers make epistemic authority visible through active triangulation between institutional data, storm chaser visuals, and audience reports. In practice, triangulation operated as a continuous "alignment" activity: creators monitored official indicators, treated chaser footage and traffic cameras as visual confirmation, and used chat as a distributed sensor network that could verify impacts faster and with more confidence than any single data feed. For example, Creator B toggled between radar and chaser context while navigating interpretive uncertainty: \textit{"[storm chaser] is on the storm, I think near Trumann, guys, this is now down to an observed [warning]. Bottom right of your screen. I'm not seeing anything completely consolidated right now."} In parallel, Creator C escalated concern when a chaser’s report (\textit{"Multiple power flashes!"}) coincided with on-screen rotation, while Creator A incorporated audience ground-truth such as \textit{"18 wheeler just turned over on highway 82 near Seymour, TX just now."} These moments show how "authority" was made visible through the integration of multiple information streams, especially during PDS periods when audiences were actively searching for confirmation.

This pattern reflects distributed cognition in the sense that interpretive work is spread across people and tools rather than centralized in a single expert source \cite{hutchins1995cognition}. Importantly, triangulation here is not merely presenting other sources, but a connective, interactive practice that displays how conclusions are reached, what is being trusted, what is provisional, and what is being cross-validated in real time. These practices contribute to performative legitimacy by making creators’ analytic reasoning publicly visible and continuously accountable during unfolding events.

\subsubsection{Audience Engagement in Information Gathering}
Across streams, creators did not treat the audience as passive recipients. Instead, they explicitly positioned viewers as contributors to a distributed observation network, soliciting local reports and then re-integrating those reports into live interpretation. Creator A verbally invited viewers to actively support one another, \textit{"Those of you watching.. Now you know, you can help answer questions via the chat. Tell them what's going to be happening."} Creator B similarly asked for reports through associated community channels: \textit{"Do we have any reports, er, people on discord or anything, just throw anything in the chats and I'll try to relay it."} These prompts were answered in-chat with highly specific ground reports (hail size, wind intensity, outages), such as \textit{"Quarter-size hail here"} or \textit{"our power is off here in Electra, TX,"} which functioned as rapid "field verification" for radar-indicated severity.

This participation extended beyond reporting current conditions. Chat members frequently answered location-based questions before the creator could, effectively performing "peer-to-peer wayfinding" or confirming where a town sits in relation to a larger metro region, or whether a location is within the current threat area. Moderators also played a visible role in maintaining safety-oriented norms by reiterating shelter guidance and occasionally redirecting viewers to local official sources when the creator’s attention was concentrated elsewhere. These behaviors align with participatory sensemaking \cite{de_jaegher_participatory_2007} and crisis-time collaborative work \cite{fischer2015building}, but the livestream context makes the mechanism unusually explicit: audience contributions are publicly evaluated, selected, and folded into the evolving account of the event.

\subsubsection{Technical Knowledge Translation}
Weatherfluencers repeatedly translated meteorological terminology and radar conventions into accessible language, enabling broader participation in interpretation while keeping the stream oriented to action. Translation ranged from defining technical terminology to embedding safety guidance within those definitions. Creator C introduced the label directly: \textit{"If you're new and you don't know what a PDS tornado warning is, it's a particularly dangerous situation..."} Creator B went further, explaining what "PDS" implies and immediately pairing it with protective action instructions. Creators also used informal analogies to demystify technical views (e.g., Creator A referring to velocity as \textit{"the which way the winds are blowing mode"}), which allowed novices to track key indicators without adopting specialist vocabulary.

Notably, translation was also socially regulated. When technical debates or corrections threatened to derail urgent communication, moderators and chat participants often enforced norms that prioritized actionable interpretation over precision disputes. For example, reminders by both moderators and participants that pronunciation corrections were unwelcome or warnings that off-topic corrections would be removed occurred regularly. In effect, translation here is a coordinated practice for stabilizing shared meaning under time pressure, another form of distributing technical cognition across expert/lay boundaries \cite{hutchins1995cognition}.

\subsubsection{Distributed Warning Network}
During high-risk periods, the livechat functioned as a secondary alert channel that amplified official warnings and circulated protective action cues. Viewers posted real-time indicators of warning activation (sirens, local alerts, town-specific warning mentions), often in rapid succession, producing a distributed chorus of warnings that reinforced urgency. Alongside these alert posts, viewers shared situated accounts of taking shelter such as reporting being in a public shelter or improvising head protection, which provided concrete, socially legible and sometimes humorous examples of actions during moments when some viewers were still asking whether the threat was serious.

Rather than treating these as mere quips, creators frequently used chat activity to calibrate attention and verify impacts, especially when multiple threats were active simultaneously or radar evidence was insufficient to interpret the storm. The result is that warning information was socially reiterated, geographically extended, and made behaviorally concrete through the visible actions of others.

\subsubsection{Transparent Analytical Process}
A final, recurring practice was the public performance of reasoning. Creators regularly narrated what they were looking at, what they could not yet confirm, and how they were weighing competing indicators. Creator B, for example, explicitly articulated uncertainty when assessing possible debris signatures and hail contamination, while still framing the implications for risk: \textit{"Latest tornado warning still showing a strong area of rotation over the east side of Bakersfield, this might be a tornado debris signature, it's also contaminated with some level of hail so cannot fully verify it, but rotation again is still on the strong side of things, we may have a tornado here near Bakersfield."} Creator A walked viewers through the interpretive logic of velocity couplets (areas of opposing winds), including numeric detail, in a way that simultaneously supported immediate understanding and modeled how their conclusions were formed. Creator C balanced transparency with audience reassurance: \textit{"I don't want to incite any panic...but I wouldn't be doing my job here if I didn't express some enhanced concern about this storm specifically."}

Chat responses during these times suggest that audiences were actively tracking and appropriating this reasoning style. Viewers echoed analytic language and sometimes challenged interpretations, indicating that transparency did not eliminate disagreement; however, it did provide shared reference points for what counts as evidence and how to interpret evolving signals.

\subsection{Temporal Sensemaking During Severe Weather Events}
Weatherfluencers engaged in temporal sensemaking by continually linking what just happened, what is happening now, and what might happen next. Across creators, these temporal strategies became more explicit during critical safety moments, when audiences faced both information volatility and practical decision pressure. We observed five recurring strategies: referencing past events, shifting into present-focused threat talk, producing anticipatory timelines, integrating multiple time horizons within single interpretive segments, and explicitly acknowledging timing limitations.

\subsubsection{Referencing Past Events}
Creators frequently used past events as comparative anchors to help audiences calibrate the present. These retrospective references served multiple communicative functions, including establishing severity baselines, contextualizing new or unusual patterns, building continuity in audience understanding, and reinforcing a shared frame of reference that supported community cohesion. Creator B, for instance, contextualized a PDS warning through recency/rarity, \textit{"the last time we had a PDS tornado warning was March 15...these aren't that common,"} contextualizing the rarity of the current situation while implicitly signaling its significance. Creator C drew on personal continuity across repeated coverage days, \textit{"This has just been the wildest roller coaster, it feels like Groundhog Day. Every day...And I'm talking about the same places every time, it's like I am reliving the same day over and over again,"} framing the unfolding period as unusually persistent. Creators also used record-breaking references, such as the number of simultaneous warnings, to signal unusual severity. These temporal-severity markers translated scale into an intelligible "this is not normal" signal for viewers.

Chat contributions reinforced this collective temporal framing through comparisons to prior outbreaks (e.g., \textit{"this is like April 12, 2020"}) and place-based memories (e.g., recalling prior local impacts). In Weickian terms, these retrospective comparisons provided ready-made interpretive resources, enabling audiences to "make the present sensible" by mapping it onto prior experience \cite{weick1995sensemaking}.

\subsubsection{Immediate Threat Communication}
When storms appeared imminent, creators compressed their temporal horizon into the present and shifted into urgent, action-oriented talk. This shift was audible and structural: shorter utterances, repeated place names, intensified imperative language, and rapid cycling between radar and confirmation sources. Creator B's phrasing illustrates how uncertainty was managed without softening urgency, \textit{"this might not be a very big tornado, but it's a strong tornado,"} and how rhetoric escalated when corroborating indicators appeared such as radar debris signatures. Creator A similarly blended immediate description with live visual context from storm chaser positions and circulation behavior, while Creator C often anchored immediacy in visual confirmation: \textit{"…there it is… producing a tornado RIGHT NOW!"}

Audience messages mirrored and amplified this immediacy through emotional reactions and directive prompts to others to seek shelter. At the same time, chat participants sometimes verified or challenged creator interpretation, indicating that immediate threat communication remains a negotiated interpretive space even at peak urgency. This pattern is consistent with temporal trajectory alignment under time pressure, where distributed actors must synchronize understanding quickly to support action \cite{reddy_temporality_2006}.

\subsubsection{Anticipatory Communication}
Outside the most immediate moments, creators invested substantial effort in translating radar patterns into actionable future storm timelines. Anticipation differed by creator: Creator A often emphasized method and timing ("We're going to put a track\footnote{A track typically means that the creator uses meteorological software with modeling capabilities to estimate a storm's path and time of arrival for a particular location, projected out for minutes to hours. This works similarly to how other storm systems, such as hurricanes, are tracked.} on this at 35mph for an hour..."), making projection logic explicit; Creator B used countdown framings tied to landmarks (\textit{"T-minus three minutes, probably…"}), coupling urgency with probabilistic hedging; Creator C frequently paired projected concern with institutional boundary limitations, warning that storms can outrun or diverge from polygons. Across creators, audience questions ("how’s X looking?") show that viewers actively used these projections for individualized risk assessment rather than general event comprehension.

Across all three streams, anticipatory communication functioned not only to forecast storm movement, but to \textit{stage decision-relevant futures} at different temporal resolutions. Creators routinely shifted between near-term countdowns, short-horizon projections (e.g., 30–60 minutes), and longer-range "later tonight" assessments, allowing viewers to calibrate urgency and preparedness over time rather than treating risk as a binary present/absent condition. By making uncertainty explicit, weatherfluencers framed anticipation as a provisional, revisable process, inviting audiences to remain attentive as conditions evolved rather than expecting definitive predictions.

\subsubsection{Temporal Integration}
Creators frequently combined multiple timeframes within single communication segments, creating "temporal bridges" that connect past, present, and future. This integration intensified during PDS warnings as creators worked to help audiences comprehend rapidly evolving threats within meaningful contexts. Creator A's use of short-term loops (\textit{"...I'll show you a loop of this over the last 20 minutes or so, and again you'll see how the first hook occluded or wrapped into the storm and now we have this new hook that is particularly concerning"}) linked prior storm structure to current risk interpretation, while Creator B’s comparison of debris lofting height connected current severity to earlier storms in the same stream. Creators also used compressed future-talk to project intensification (\textit{"This is a significant tornado. This is going to wedge."}\footnotemark), tying present indicators to near-future consequences.
\footnotetext{"Going to wedge" refers to a tornado growing so large that it appears as a triangular-shaped wedge emerging from the sky. This is typically thought of as one of the most destructive types of tornado, although any can produce significant damage.}

Creator B similarly integrated multiple timeframes when contextualizing debris height: \textit{"Also, debris from this tornado and the tornado emergency is being lofted 31,000 feet into the atmosphere. This is a substantial tornado. For reference, our tornado that happened earlier, about 3 hours back...that one was lifting debris up about 12,000 feet. That tornado is not nearly as significant as what we're seeing with this tornado right now."}

Importantly, we observed a consistent constriction of temporal scope at peak threat: during the most severe confirmed moments, creators and chat tended to collapse into the immediate present (what is happening now, where to shelter now), temporarily deprioritizing broader contextualization. This suggests temporal integration is not constant, but dynamically modulated as urgency narrows the communicative task from understanding to immediate action.

\subsubsection{Explicit Timing Limitations}
Finally, creators frequently acknowledged timing uncertainty as a feature of the information environment. These acknowledgments included lags in spotter reports, mismatches between notification systems and on-screen tools, and overload when multiple warnings arrived close together. Creator B demonstrated this transparency when noting, \textit{"there's a report in [town]... tornado has been reported, this is a PDS tornado, I don't know if I missed it, but there was a large stove pipe tornado reported approximately 10 minutes ago heading towards Trumann."} This explicit acknowledgment of information lag and information overload highlights the challenges of real-time reporting while maintaining credibility through transparency. Creator B also addressed timing uncertainty in radar updates: \textit{"My system isn’t slow, by the way, our tornado warning notifications come in faster a lot of the time than Radar[Scope]’s do, I’m just letting you guys know like a warning will come in soon and it just takes a minute or whatever."} This meta-commentary about notification systems helps audiences understand the potential mismatch between official warnings and what they are seeing on screen. Across creators, these moments operated as credibility maintenance and made the constraints of real-time multi-source interpretation visible, particularly during high-volume warning periods.

\subsection{Platform Affordances and Constraints in Safety-Critical Communication}

Weatherfluencers adapted YouTube’s entertainment-oriented infrastructure into safety-critical communication systems through four recurring techniques: multi-window integration, visual annotation, technical workarounds during breakdowns, and attention management across simultaneous threats. Across creators, the work of "making YouTube usable for crisis communication" was both technical (tool switching, layout composition) and interactional (norm-setting, triage, continuity signaling).

\subsubsection{Multi-Window Visual Presentations}
\begin{figure}[ht]
  \centering
\includegraphics[width=0.8\linewidth]{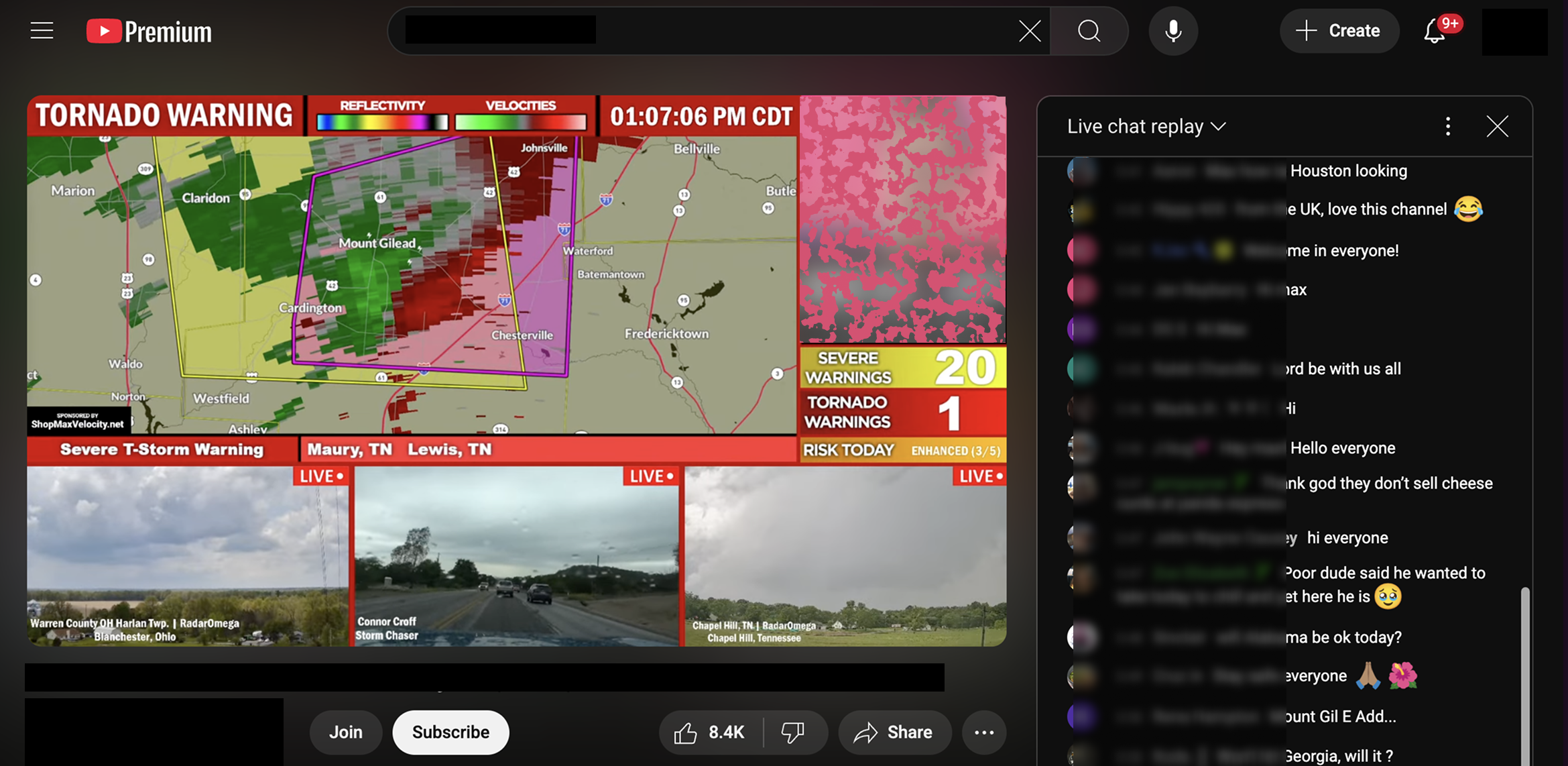}
  \caption{Example of a livestream with YouTube chat interface}
  \Description{Image showing weatherfluencer user interface to communicate about radar indicated storms with various storm warning metrics and live chat on right rail}
  \label{fig:yt_interface_chat}
\end{figure}

All three creators relied on multi-window layouts to coordinate heterogeneous sources such as radar, warning maps/overlays, traffic cameras, and chat so that audiences could view evidence and its interpretation in the same visual field. Creator A narrated across windows while tracking chaser position relative to circulation; Creator B frequently switched among multiple intercept feeds to confirm evolving reports; Creator C integrated public cameras to render impacts legible under heavy rain/wind. During PDS warnings, multi-window complexity typically increased (often 2–4 simultaneous visual elements), accompanied by overlays that maintained a running account of active warnings even when the creator’s spoken attention was elsewhere. Figure \ref{fig:yt_interface_chat} shows an example of this multi-window configuration, illustrating how creators visually coordinate diverse information sources while engaging with live chat.

Creator C described this approach, \textit{"What you're looking at on these traffic cameras is some of the strong winds and heavy rain moving into the areas that are getting overtaken by the bow echo that we're watching [on radar]."}  Rather than treating this as presentation polish, the key finding is that multi-window configurations acted as interpretive scaffold: they enabled rapid cross-referencing under uncertainty and made the evidentiary basis of claims publicly inspectable. This is consistent with infrastructuring work as the ongoing adaptation of a system to meet emergent needs \cite{star_infrastructure_2002}, but here, the infrastructural "repair" is performed live, on-stream, under time pressure.

\subsubsection{Visual Annotation Techniques}
\begin{figure}[ht]
  \centering
\includegraphics[width=0.8\linewidth]{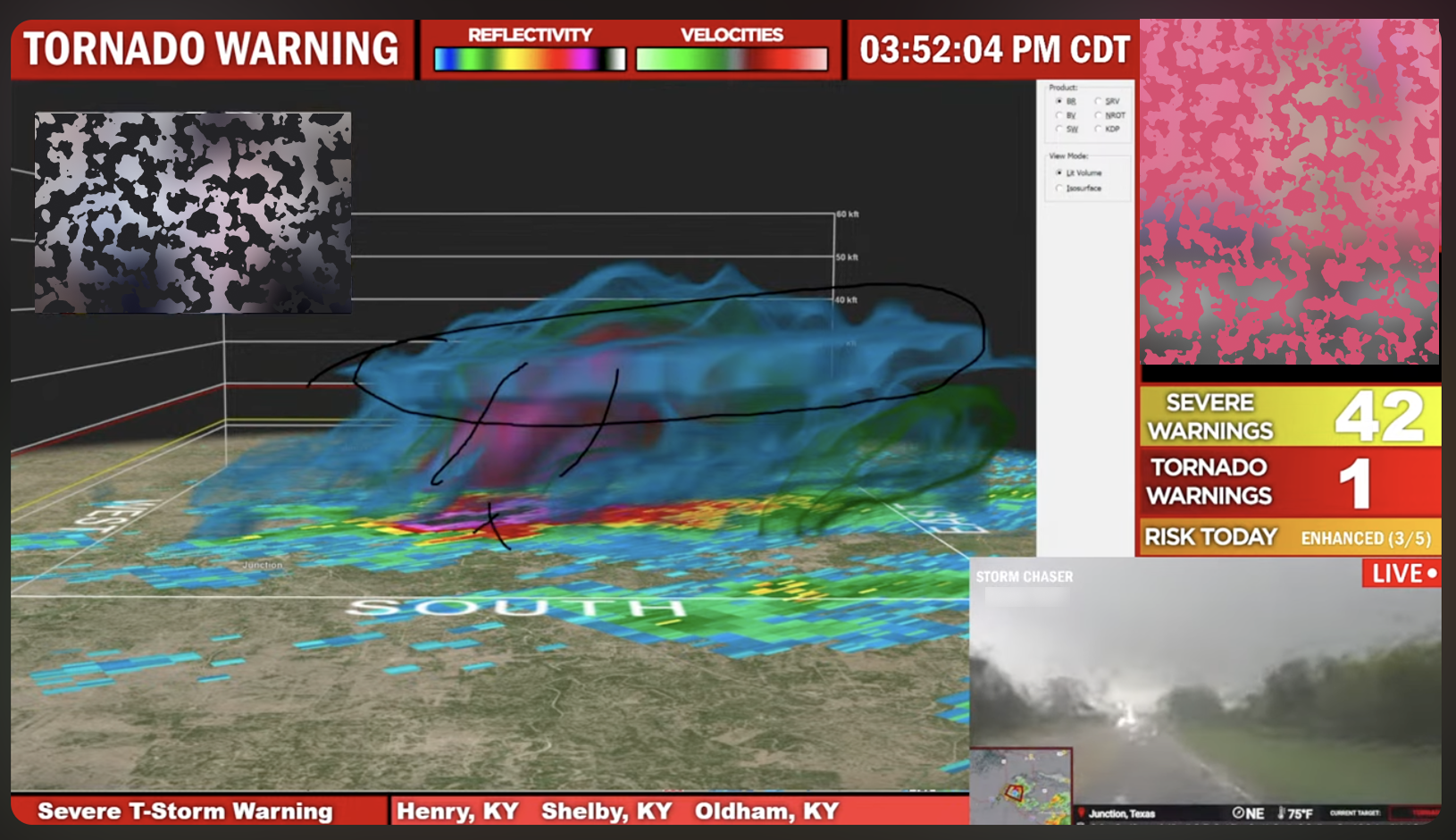}
  \caption{Example of a livestream with illustration-assisted explanations of storm structures}
  \Description{Image showing weatherfluencer user interface and pen tool to hand-draw explanations on top of an image of a storm structure.}
  \label{fig:pen_tool}
\end{figure}

Creators used digital pen tools and other overlays to render analytic reasoning visible, for example, circling rotation, tracing projected movement, and marking locations, especially when communicating projected paths or escalating concern. Creator B’s annotation practices often emphasized evidentiary emphasis (e.g., highlighting debris signatures and motion), while Creator A used markings to localize risk (\textit{"I’ll put it on the map…"}) and help viewers orient geographically. These annotations were not constant; they appeared most frequently when creators needed to translate a complex spatial-temporal claim into a quickly graspable visual story. Figure \ref{fig:pen_tool} illustrates these annotation practices.

Chat engagement frequently referenced these visuals (asking what a feature was, reacting to visible lightning or precipitation), suggesting that annotation helped synchronize attention and provided a shared object for questions during rapid updates.

\subsubsection{Technical Challenges and Adaptations}
Because creators relied on third-party radar tools and stable network conditions, breakdowns were common and the key empirical point is how creators and audiences jointly managed them. Creators openly narrated tool switching and improvised alternatives (e.g., moving between radar applications), often using humor or candid commentary to maintain continuity (\textit{"the radar hates you, but I'll show you a different source"}). Creator C, in contrast, sometimes framed limits in precision as a methodological constraint, prioritizing real-time video or confirmed reports when radar ambiguity was high: \textit{"I don't want to be too precise [about the storm location], but if we can be extremely precise, the way to do that is with real-time video and not just through radar."}

Audience members actively contributed to continuity during disruptions by answering tool questions, prompting refresh behavior, and signaling when streams returned. When connectivity dropped, chat often became a temporary coordination layer (\textit{"good here and [channel] live is back."} ), while viewers simultaneously posted local impact updates (sirens, outages), keeping a minimal situational picture alive even when the creator’s feed degraded.

\subsubsection{Attention Management Across Multiple Threats}
During widespread outbreaks, creators adopted explicit triage frameworks: focusing on the most severe storm, prioritizing areas with higher viewer concentration, and using on-screen elements to maintain peripheral awareness of other warnings. Creator B noted this challenge explicitly: \textit{"We have tornado warnings again everywhere, it's going to be impossible for me to cover every single tornado warning."} Creator C used geographic prioritization to manage attention, focusing on areas with viewer concentrations: \textit{"For our friends in TN, I know we had a bunch of people watching from TN because we had a crazy tornado situation of our own up there not too long ago, the severe weather threat is diminishing quickly for the vast majority of Tennessee,"} later using "release talk" (\textit{"yes, you can go to bed now"}) to help subsets of the audience disengage safely as threat diminished. Creator A reassured those outside the focus area that they likely had nothing to worry about when the stream's attention went elsewhere, \textit{"Folks, if I'm not talking about where you are right now or talking about anywhere near you right now, you have nothing to worry about!"}

Chat norms reinforced this triage by discouraging repeated individualized location questions and redirecting viewers toward local sources when appropriate: \textit{"stop asking about your area, turn on your local news for that."} Supportive messages also indexed recognition of the labor of triage (thanking creators, noting stress or "freaked out" moments), which appeared to function as social maintenance during long-duration high-load periods. Across streams, visual banners/overlays increased with the number of active warnings, helping audiences self-select relevance while the creator maintained focus on the most acute threat.

Across these findings, we understand creators' authority as emerging not primarily from formal credentials, but as enacted through ongoing, visible interpretive work, what we describe as performative legitimacy, as creators triangulate sources, project timelines, and reason publicly under uncertainty. While these practices are often associated with credibility in prior work, our analysis does not evaluate their effects on audience trust, but instead examines how they are mobilized within real-time interpretation.

\section{Discussion}
Livestreamed weather coverage can look like a simple extension of citizen science or participatory media. Critical analysis reveals it to be something far more transformative: a site where epistemic authority, institutional trust, and civic safety are publicly negotiated in real time. Weatherfluencers' practices unsettle existing theories of crisis informatics, sensemaking, and platform studies, while revealing urgent sociopolitical conditions shaping their emergence. Weatherfluencers are not simply content creators but can be understood as emerging forms of informal epistemic infrastructure whose authority is shaped by technical skill, interactional labor, and institutional voids.

\subsection{The Emergence of Networked Crisis Expertise}
In traditional crisis communication models, expertise is institutional, formalized, and top-down. In participatory models, it is crowdsourced, fragmented, bottom-up, and episodic. The practices observed in severe weather livestreams align fully with neither. Instead, weatherfluencers enact a form of networked crisis expertise in which authority is made visible through the real-time synthesis of data, social interaction, visual narration, and responsiveness under conditions of uncertainty.

This expertise is both situated in the creator and distributed across a sociotechnical assemblage through the chat contributors offering hyperlocal updates, moderators regulating information flow, storm chasers providing live on-the-ground visuals, and institutional sources (National Weather Service alerts, radar images) being interpreted on the fly. The creator mediates these flows, legitimizing some signals, discounting others, and narrating uncertainty as it evolves. This sustained responsibility for public interpretation distinguishes weatherfluencer practice from more episodic forms of digital participation.

These dynamics extend Hutchins’ \cite{hutchins1995cognition} model of distributed cognition by illustrating how collective interpretation occurs in volatile, public-facing environments without stable team membership, shared training, or institutional coordination. Unlike the bounded teams Hutchins examined, participants in these livestreams must co-construct meaning across heterogeneous expertise levels, shifting attention, and platform-mediated constraints. Authority in this setting is therefore not embedded in role or credential alone, but enacted through ongoing interpretive work that remains visible to audiences and carries reputational risk.

At the same time, weatherfluencer sensemaking challenges Weick’s emphasis on retrospective coherence. Rather than constructing meaning after outcomes are known, creators operate in a prospective mode in which interpretation must be articulated in advance of consequences. Creators routinely reason aloud about hypothetical futures, projecting storm paths, estimating arrival times, and qualifying uncertainty, because delayed interpretation may carry life-threatening implications. Sensemaking here is oriented less toward explanation than toward timely orientation and action under uncertainty.

Across these practices, networked crisis expertise becomes credible through visible and audible interpretive work, suggesting that legitimacy is enacted interactionally rather than conferred in advance. We return to the implications of this performative legitimacy for trust and credibility later in Section \ref{trust_cred_auth}.

\subsection{Livestreaming as Public Temporal Sensemaking Infrastructure}
Traditional sensemaking frameworks, particularly those drawing from Weick, emphasize retrospective interpretation, where meaning is constructed after events have occurred \cite{weick1995sensemaking}. In crisis contexts, this retrospective orientation is challenged by the need for real-time response and anticipatory actions. Weatherfluencers operate at a unique temporal intersection where these distinct modes of sensemaking, retrospective, real-time, and anticipatory, collapse into a synchronous communicative practice.

Unlike institutional crisis communication that often separates historical analysis from current status updates and future projections, livestreams and the nature of the weather domain create what we term "\textit{public temporal sensemaking infrastructure}": sociotechnical arrangements that enable distributed audiences to navigate multiple timeframes simultaneously. This infrastructure is also deeply relational \cite{star_steps_1994}, emerging through the dynamic between creators, audiences, and platform affordances during high-stakes events. 

The public nature of this sensemaking process distinguishes it from the team-based temporal coordination observed in organizational settings \cite{reddy_temporality_2006, weick1995sensemaking}. On platforms designed for entertainment rather than emergency response, weatherfluencers engage in ongoing infrastructuring work \cite{pipek_infrastructuring_2009}, adapting existing systems to meet emergent needs. Through visual annotations, multi-window displays with remote storm chasers, and narrative riding techniques, they transform platforms into interpretive tools that make temporal relationships explicitly visible and tightly condensed. 

Rather than becoming visible only upon infrastructural breakdown, as Star observed \cite{star_infrastructure_2002}, weatherfluencer practices proactively expose the temporal architecture of weather. By linking past patterns, real-time observations, and anticipated threats into continuous narrative form, creators help viewers situate themselves in both space and time. This temporal scaffolding becomes especially vital during moments of heightened uncertainty, when institutional communications may struggle to bridge diverse temporal and informational needs.
 
\subsection{Platform Appropriation and Civic Infrastructure}
Our findings reveal how platform affordances and constraints fundamentally shape weatherfluencers' communication practices, often in ways that transform entertainment platforms into improvised civic infrastructure. The multi-window visual presentations, visual annotation techniques, and attention management strategies we documented represent systematic adaptations that leverage YouTube's technical capabilities while working around its limitations.

Weatherfluencers transform YouTube's entertainment-oriented features into safety-critical communication tools. The multi-window layouts that enable simultaneous radar data, storm chaser footage, and chat interaction reflect what Star and Pipek \cite{star_infrastructure_2002, pipek_infrastructuring_2009} describe as infrastructuring work: the ongoing adaptation of existing systems to serve emergent needs. Through these configurations, creators appropriate platform affordances beyond their intended design \cite{dourish2003appropriation}, enabling real-time triangulation across heterogeneous information sources. These arrangements fundamentally structure how interpretation unfolds under time pressure.

The technical challenges we observed reveal the precarious nature of using entertainment platforms for emergency communication. Breakdowns in third-party tools, streaming latency, and data outages are treated not as exceptional events, but as routine contingencies that creators must anticipate and manage in real time. When Creator B switches radar systems mid-stream, it reveals how platform-mediated crisis communication requires continuous improvisation to maintain interpretive continuity, often under conditions where failure could have serious safety implications.

Platform constraints also redistribute responsibility for coordination and prioritization. Limited geographic targeting and chat-driven issue surfacing force creators to make explicit triage decisions about which warnings to follow, which locations to emphasize, and which signals to ignore. Chat communities often participate in this work by flagging locations, requesting clarification, or reminding others of established norms, effectively sharing responsibility for managing attention under constraint. These decisions shape what information becomes visible and actionable for distributed audiences. 

These dynamics unfold within the broader logic of social media platforms, where programmability, popularity, and connectivity privilege engagement over reliability \cite{van2013understanding}. As a result, weatherfluencers and their audiences must compensate for platform misalignment by developing informal coordination practices that approximate civic infrastructure without institutional support. In doing so, they expose both the possibilities and the limits of repurposing entertainment platforms for safety-critical communication.

\subsection{Trust, Credibility, and Authenticity in Collaborative Crisis Communication}
\label{trust_cred_auth}
Rather than evaluating whether these practices result in audience trust or credibility, we examine how credibility is made visible and actionable within interaction, without assessing whether audiences ultimately perceive these practices as credible or trustworthy.

While trust was not posed as a standalone research question, it emerged as a central implication of the collaborative sensemaking practices (RQ1), temporal coordination (RQ2), and platform constraints (RQ3) documented above. Trust fundamentally shapes whether people take protective action during severe weather events \cite{ripberger_false_2015, sherman-morris_perceived_2020, trainor_tornadoes_2015}, yet crisis informatics has often examined credibility as an outcome assessed retrospectively or relative to institutional benchmarks. Our findings instead foreground credibility as something that must be \textit{made visible and actionable in real time} to sustain collective sensemaking under conditions of uncertainty.

Across severe weather livestreams, we understand these practices as enacting what we term \textit{performative legitimacy}: credibility rendered visible through ongoing interpretive labor rather than conferred by institutional position alone. This legitimacy is enacted through practices such as triangulating multiple data sources, reasoning aloud about uncertain futures, annotating radar imagery, and responding to audience questions as events unfold. While some creators possess formal meteorological credentials, their authority in these moments appears to rest less on certification than on visible analytical competence, narrative coherence, and affective attunement during moments of shared vulnerability. Performative legitimacy is provisional and contingent, but it is interactionally sufficient to keep sensemaking moving when delay or silence could carry serious consequences.

This analysis suggests a shift in how credibility can be understood in crisis contexts. Rather than being evaluated and examined after the fact, credibility in livestreamed weather communication appears to be continually tested and repaired in public view. Verification processes that remain largely invisible in broadcast meteorology are made explicit, allowing audiences to observe not only conclusions but the reasoning and uncertainty management that underpin them. In this setting, credibility can be understood as distributed across creators, audiences, and tools, while remaining anchored in the creator's visible interpretive responsibility.

However, platform affordances make this form of credibility fragile. Technical failures, algorithmic mediation (how viewers become aware of the channels), and limited geographic targeting introduce new vulnerabilities that do not exist in traditional broadcast environments. Creator communications consistently acknowledged analytical limitations and uncertainties, reflecting established practices from broadcast meteorology. Traditional TV meteorologists similarly avoid absolute statements to prevent panic and maintain credibility over time \cite{sherman-morris_perceived_2020}. 

Maintaining credibility under these conditions often becomes a collaborative achievement, with chat participants assisting in information repair, technical troubleshooting, and attention management, forms of distributed livestream labor adjacent to volunteer moderation practices documented on Twitch \cite{seering_moderates_2022}. Rather than representing entirely new credibility mechanisms, these patterns suggest how established meteorological communication practices adapt to platform affordances that enable continuous, visible uncertainty management rather than periodic authoritative statements. The collaborative chat environment allows audiences to observe analytical work in real-time, potentially making traditional credibility-building techniques more visible through increased transparency. These dynamics highlight how platform design choices shape not only information visibility, but the stability of trust itself during safety-critical moments.

While the traditional relationship between credibility, credentials, and trust is well-established in weather communication research \cite{ripberger_false_2015, sherman-morris_perceived_2020, trainor_tornadoes_2015}, the weatherfluencer phenomenon appears to be supplemented by alternative credibility pathways that operate through transparency, collaborative verification, and visible analytical labor. Rather than viewing informal communicators as supplements to official sources, our findings suggest they may represent an evolution in how crisis communication is organized to meet public needs during geographically distributed, time-sensitive threats.

\subsection{Platform Design Implications}
Our findings show how platforms designed for entertainment are repeatedly co-opted for safety-critical purposes, a pattern long observed in crisis contexts. Weatherfluencer livestreams exemplify this misalignment: platforms optimized for engagement and scale are repurposed as ad hoc civic infrastructure, forcing creators and audiences to compensate for gaps in reliability, geographic specificity, and situational clarity. Viewers arrive at these streams for information, but also to locate themselves physically, temporally, and affectively within unfolding threats. Yet platforms largely treat specialized livestreams as one-size-fits-all, offering little support for the types of affordances weatherfluencers have evolved to develop in their proprietary user interfaces.

A first implication concerns geographic relevance. Chat interactions frequently revealed tension between local and non-local viewers, as participants asked creators to focus on specific locations, disrupting chat flow and prompting moderators to reassert norms. Platforms currently offer no support for resolving this tension. Large language models (LLMs) could potentially function as filtering infrastructure, helping route viewer attention toward stream segments likely relevant to their location based on ongoing discourse and visual context, without automating interpretation or displacing creator judgment.

Second, temporal scaffolding remains fragile. While weatherfluencers attempt to maintain continuity through repetition, pinned messages, and visual annotation, these efforts are easily lost amid rapid chat flow and interface clutter. Structured safety templates, such as persistent panels summarizing active warnings, projected timelines, and uncertainty ranges, could help stabilize "what is currently known" without requiring creators to continually restate context. However, these designs must be carefully constrained: overly prominent alerts risk amplifying panic or spectacle, particularly given algorithmic incentives toward drama.

Third, infrastructural reliability poses a persistent challenge. The frequent technical breakdowns observed in our data underscore creators’ reliance on third-party tools and improvised workarounds during safety-critical moments. Rather than novel systems, platforms could support crisis communication through basic redundancy mechanisms, such as preconfigured backup visualizations, cached data states, or rapid tool switching, to reduce disruption when primary systems fail.

Finally, credibility signaling warrants careful design attention. While legitimacy in weatherfluencer streams is earned interactionally, the absence of standardized cues leaves viewers to infer expertise through complex, creator-built interfaces that often resemble over-engineered dashboards rather than usable safety tools. Platform-level support for lightweight credibility signals, such as visibility for the American Meteorological Society’s Certified Digital Meteorologist (CDM) designation \cite{AMSCertified2025} could provide optional, interpretable indicators of expertise without overriding the performative legitimacy that anchors these communities. Additional signals, such as transparent data sourcing or update latency, could further support informed trust.

In the absence of sufficient institutional presence and amid the opacity of platforms, weatherfluencers provide not just information, but orientation. They help distributed publics locate themselves physically, temporally, and affectively within unfolding threats. This orienting work extends beyond purely informational or infrastructural functions and becomes increasingly critical as public safety communication grows more decentralized. Designing platforms that support such orienting practices will therefore be essential for the future of crisis informatics and emergency management.

\section{Limitations}
Our study provides insight into collaborative sensemaking during severe weather events, but several limitations should be considered. First, our analysis focused on only three YouTube creators, which cannot capture the full diversity of practices in the weatherfluencer ecosystem. This small sample, primarily consisting of established creators, limits our ability to identify variations across demographics, expertise levels, and audience sizes. Additionally, while this analysis documents how creators structure and perform sensemaking with their audiences, it does not assess how viewers definitively interpret, trust, or act on the information presented. As a result, the findings speak to the availability and organization of participatory sensemaking within livestreams, rather than to the effectiveness or accuracy of these practices from the audience's perspective.

Second, our critical incident sampling approach, while effective for examining high-stakes communication, may overlook important patterns during less urgent periods that establish baseline knowledge and audience relationships. Our analysis captures observable communication practices without accessing creator intentions or audience interpretations, limiting our understanding of the effectiveness of these communications from both perspectives.

Third, our focus on tornado outbreak events in the United States during Spring 2025 limits generalizability to other crisis types, cultural contexts, and time periods. Different hazards might elicit different temporal orientations and collaborative dynamics. Future research should expand to more diverse creators across platforms, incorporate audience perspectives through surveys or interviews, compare communication across hazard types, and further investigate how trust and credibility are constructed in these contexts in a livestream environment.

Lastly, we emphasize that this study does not assess persuasion, belief, or downstream behavioral outcomes. Instead, it focuses on how collaborative sensemaking is made possible through livestream affordances and observable audience participation.

\section{Conclusion}
We examined weatherfluencer livestreams during Particularly Dangerous Situation (PDS) warnings to understand collaborative sensemaking in severe weather communication. We show how weatherfluencers orchestrate collective interpretation by triangulating multiple data sources, engaging audiences as contributors, and bridging past observations, present conditions, and anticipated futures.

Our findings suggest that weatherfluencers function as interpretive intermediaries between institutional warning systems and distributed publics. Rather than relying on institutional credentials alone, their authority is treated as emerging through visible analytical work: reasoning aloud, managing uncertainty, and responding to audience input in real time. 

This work reveals tensions that arise when crisis communication occurs on platforms optimized for engagement rather than reliability or situational clarity. By foregrounding real-time, multimodal sensemaking, this study extends research in both computer-supported cooperative work and crisis informatics beyond retrospective and 
text-based settings, highlighting the growing importance of platform-mediated interpretive work in contemporary crisis communication.

\bibliographystyle{ACM-Reference-Format}
\bibliography{MAIN}

\appendix
\section{Appendix: PDS Warning Analysis Template}
\label{appendix:analysis-template}
This appendix provides the systematic analysis template used to code weatherfluencer livestreams and live chats during Particularly Dangerous Situation (PDS) warnings.

\subsection*{Stream Information}
\begin{itemize}
    \item \textbf{Creator/Channel:} 
    \item \textbf{Date/Event:} 
    \item \textbf{Stream Link:} 
    \item \textbf{Stream Length:} 
\end{itemize}

\subsection*{PDS Warning Overview}
\begin{center}
\footnotesize
\setlength{\tabcolsep}{3pt}
\begin{tabular}{|c|c|c|c|c|c|c|c|}
\hline
\textbf{Warning\#} & \textbf{Preview Time} & \textbf{Start Time} & \textbf{End Time} & \textbf{PostView Time} & \textbf{Duration} & \textbf{Location} & \textbf{Notes} \\
\hline
 &  &  &  &  &  &  &  \\
\hline
\end{tabular}
\end{center}

\subsection*{Individual PDS Warning Analysis}
\subsubsection*{PDS Warning \#\_\_}
\begin{center}
\footnotesize
\setlength{\tabcolsep}{3pt}
\begin{tabular}{|p{2.8cm}|p{2.8cm}|p{0.8cm}|p{2.3cm}|p{1.3cm}|}
\hline
\textbf{Domain} & \textbf{Code} & \textbf{Time} & \textbf{Example/Quote} & \textbf{Notes} \\
\hline
Collective Sensemaking & Information Triangulation &  &  &  \\
\hline
 & Query-driven &  &  &  \\
\hline
Sensemaking Process & Knowledge Scaffolding &  &  &  \\
\hline
 & Transparency &  &  &  \\
\hline
Temporal Orientation & Real-time Interpretation &  &  &  \\
\hline
 & Speculation &  &  &  \\
\hline
Platform Affordances & Visual Affordances &  &  &  \\
\hline
 & Platform Limitations &  &  &  \\
\hline
\end{tabular}
\end{center}

\clearpage

\section{Codebook: Themes, Codes, Descriptions, and Examples}
\label{appendix:codebook}
\begin{center}
\footnotesize
\setlength{\tabcolsep}{3pt}
\begin{tabular}{|p{2.5cm}|p{2.5cm}|p{4cm}|p{4cm}|}
\hline
\textbf{Theme} & \textbf{Code} & \textbf{Description} & \textbf{Example} \\
\hline
Collective Sensemaking & Information Triangulation & Audience providing ground observations that confirm or challenge radar data & ``Thanks for that report from Tulsa about the hail size - that matches what the radar signature suggests'' \\
\cline{2-4}
 & Query-driven & Creator redirecting attention based on audience questions or concerns & ``Several people are asking about the storm near Oklahoma City - let me switch to that cell and explain what I'm seeing'' \\
\hline
Sensemaking & Knowledge Scaffolding & Creator building understanding by connecting technical data to familiar concepts. Often educational moments & ``See this hook echo and red-green velocity couplet? It's like looking at two gears turning against each other - when they're this tight and intense, it shows strong rotation that often precedes tornado formation'' \\
\cline{2-4}
 & Transparency & Creator verbally working through their analysis process so audience can follow along & ``I'm looking at the velocity data here, the red next to green tells me it's rotating, and combined with this debris signature, that suggests a tornado is likely forming right now'' \\
\hline
Temporality & Retrospective framing & Recalling earlier events to make sense of now & ``This is similar to what we saw in Moore earlier today.'' \\
\cline{2-4}
 & Real-time interpretation & Describing or inferring live radar activity & ``We're seeing a debris ball form *right now*.'' \\
\cline{2-4}
 & Speculation & Predicting future movement or risk & ``This storm is likely headed toward Tuscaloosa in 30 minutes.'' \\
\cline{2-4}
 & Temporal uncertainty & Expressing limits of temporal knowledge & ``It's hard to say how fast this is moving; it keeps shifting.'' \\
\hline
Platform affordances and constraints & Attention to visual affordances & Use of radar/graphics to support claims & ``Here on the left, you can see the hook echo developing.'' ``Let's pull up a traffic cam to get a better look at this storm'' \\
\cline{2-4}
 & Platform limitations & Commenting on UI, chat flow, lag & ``Looks like we lost our [storm chaser]'' \\
\hline
\end{tabular}

\vspace{0.5em}
\textbf{Table 1.} A lightweight, focused coding schema for weather communication analysis
\label{tab:weather-communication}
\end{center}

% \section{Acknowledgments}
% This section has a special environment:
% \begin{verbatim}
%   \begin{acks}
%   ...
%   \end{acks}
% \end{verbatim}

\end{document}